\documentclass[aps,prb,reprint]{revtex4-1}

\usepackage{graphicx}%
\usepackage{dcolumn}%
\usepackage{bm}%

\draft 

\begin{document}


\title{Robust luminescence of the silicon-vacancy center in diamond at high temperatures} 



\author{Stefano Lagomarsino,$^{1,2}$ Federico Gorelli,$^{3,4,2}$
Mario Santoro,$^{3,4,2}$ Nicole Fabbri,$^{3,4}$
Ahmed Hajeb,$^{4}$ Silvio Sciortino,$^{1,2}$ Lara Palla,$^{5,6}$
Caroline Czelusniak,$^{1,2}$ Mirko Massi,$^{2}$
Francesco Taccetti,$^{2}$ Lorenzo Giuntini,$^{1,2}$ Nicla Gelli,$^{2}$
Dmitry Yu Fedyanin,$^{7}$ Francesco Saverio Cataliotti,$^{1,4,8}$
Costanza Toninelli,$^{3,4,8}$ and Mario Agio$^{3,4,8,9,*}$}

\affiliation{
$^{1}$ Department of Physics and Astrophysics, University of Florence, 50019 Sesto Fiorentino, Italy\\
$^{2}$ Istituto Nazionale di Fisica Nucleare, Sezione di Firenze, 50019 Sesto Fiorentino, Italy\\
$^{3}$ National Institute of Optics (INO-CNR), 50019 Sesto Fiorentino, Italy\\
$^{4}$ European Laboratory for Nonlinear Spectroscopy (LENS), 50019 Sesto Fiorentino, Italy\\
$^{5}$ Istituto Nazionale di Fisica Nucleare, Sezione di Pisa, 56127 Pisa, Italy\\
$^{6}$ Department of Physics, University of Pisa, 56127 Pisa, Italy\\
$^{7}$ Laboratory of Nanooptics and Plasmonics, Moscow Institute of Physics and Technology, Dolgoprudny 141700, Russian Federation\\
$^{8}$ Center for Quantum Science and Technology in Arcetri (QSTAR), 50125 Florence, Italy\\
$^{9}$ Laboratory of Nano-Optics, University of Siegen, 57072 Siegen, Germany}

\email[Corresponding author. ]{mario.agio@uni-siegen.de; agio@lens.unifi.it}


\date{\today}

\begin{abstract}
We performed high-temperature luminescence studies of silicon-vacancy
color centers obtained by ion implantation in single crystal diamond.
We observed reduction of the integrated fluorescence upon increasing
temperature, ascribable to a transition channel
with an activation energy of 180 meV that populates a shelving state.
Nonetheless, the signal decreased only 50\% and 75\% with
respect to room temperature at 500 K and 700 K, respectively.
In addition, the color center is found highly photostable
at temperatures exceeding 800 K.
The luminescence of this color center is thus extremely robust even
at large temperatures and it holds promise for
novel diamond-based light-emitting devices. 
\end{abstract}


\maketitle 


\newcommand{\tab}{\hspace{5mm}}

Diamond and silicon carbide are well known for
their potential in electronics, especially for
high-speed, high-frequency, high-temperature,
and high-power applications.\cite{willander06}
Recent developments in the electroluminescence
of color centers in diamond\cite{lohrmann11,mizuoki12,berhane15} 
and silicon carbide\cite{lohrmann15}
make the assessment of their luminescence properties above room temperature
particularly interesting. In fact, at relatively high temperatures
the populations of the valence and conduction bands, which are
typically quite small due to the large
activation energy of n- and p-type dopants in diamond,
should significantly increase\cite{gabrysch08,stenger13}
and thus potentially lead to much higher electroluminescent rates.

The temperature dependence of the luminescence peaked at 738 nm of the
negatively-charged silicon-vacancy (SiV$^-$) center in diamond
has long been studied in the range between cryogenic and
room temperatures.\cite{pezzagna11,feng93,gorokhovsky95,neu13}
Those measurements revealed, along with a red shift and a widening of
the zero-phonon line (ZPL) with increasing temperatures, a quenching
of the luminescence presumably due to the thermal activation of a
non-radiative decay mode with an activation energy evaluated 
in Ref.~\onlinecite{feng93} of about 70 meV. The integrated signal seemed
to reach a plateau beyond 250 K, placed at about 10\%\cite{feng93}
or 20\%\cite{pezzagna11} of the low temperature limit.

In this Letter, we report on the luminescence properties
of SiV\textsuperscript{-} color centers in diamond
at temperatures from 293 K to 861 K.
In this range, we observed a further
reduction of the integrated luminescence ascribable 
to population transfer into a shelving state,\cite{neu12}
facilitated by a transition channel with an activation energy of 180 meV.
Nonetheless, the decrease of the signal is only 50\% of the room-temperature
level at about 500 K and at about 700 K it reaches a value 
corresponding to 25\% of that at 293 K.
The intensity is still decreasing at further higher temperatures,
although at a reduced rate.
The quenching takes place along with a red-shift of the 738 nm peak
up to 751 nm at 861 K, and with an increase of the
Full Width at Half Maximum (FWHM) from less
than 6 nm at 293 K to more than 80 nm at 861 K, while the phonon 
side-band becomes progressively less important down to complete
quenching at about 600 K.

We implanted 8 MeV Si-ions in an optical grade single crystal
diamond of $4.5 \times 4.5 \times 0.5$ mm\textsuperscript{3}
(ElementSix (R)).
Sample irradiation was carried out at the electrostatic deflector
facility (DEFEL) of the LABEC accelerator laboratory in Florence.
The ultrafast voltage transitions of the deflector allow ions
to be transmitted downstream and implanted in the sample.
The mean number of ions per bunch can be very small,
even less than one particle per bunch.
Targets are mounted inside a vacuum chamber on a frame,
where step motors drive their displacements ($\pm$ 20 cm)
with a reproducibility within few microns.\cite{palla15}

\begin{figure}[htbp]
\begin{center}
\includegraphics[width=8.5cm]{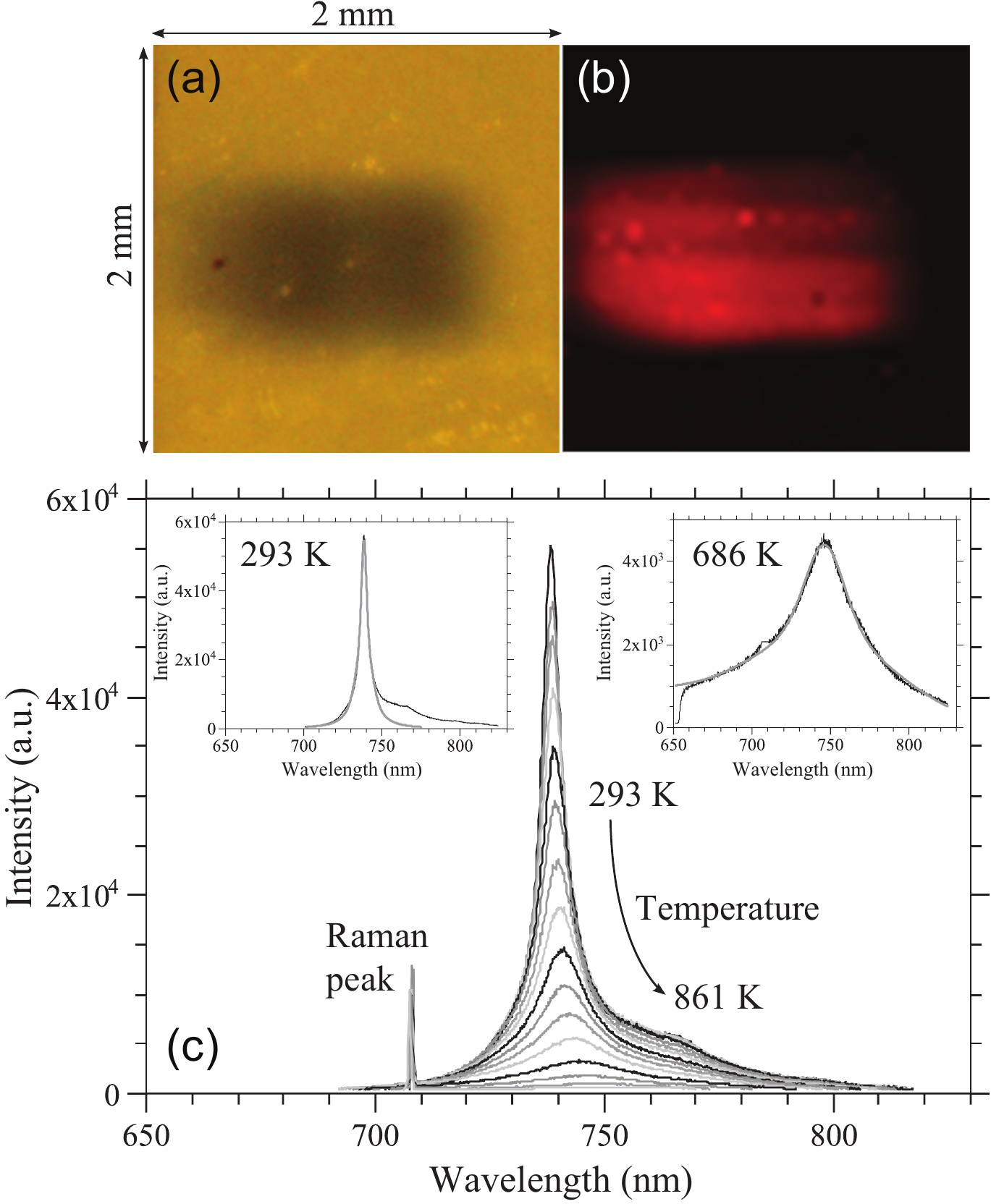}
\caption{\label{fig1} 
(a) Photographic image of the implanted area of the sample,
before temperature annealing.
(b) Luminescence map of the same region acquired at 293 K, after annealing.
(c) Luminescence spectra acquired at increasing temperatures
(the curves correspond to 293 K, 312 K, 321 K, 340 K,
356 K, 376 K, 399 K, 425 K, 454 K,
486 K, 524 K, 569 K, 623 K, 686 K, 762 K, 861 K
with decreasing peak height);
a linear background has been subtracted in order to take into account
the thermal emission at high temperatures.
Insets: fit of the ZPL at 293 K and 686 K 
with a pseudo-Voigt distribution.}
\end{center}
\end{figure}

The size of the implanted area, defined by means of two pairs of slits
placed about 50 cm before the sample along the ion beam line,
is about $1.2 \times 0.6$ mm\textsuperscript{2}, with 
an uncertainty due mainly to the divergence of the beam, which is
about 0.3 mrad. 
The irradiation fluence was evaluated to be
$6 \times 10^{14}$ cm\textsuperscript{-2}.
The implantation depth distribution, calculated with the Monte Carlo SRIM code,\cite{ziegler85}
is peaked at 2.1 \ensuremath{\mu}m, with a FWHM of 100 nm, 
thus the average volumetric density around the Bragg peak is around
$6 \times 10^{19}$ cm\textsuperscript{-3}.
After irradiation, the sample presents a strong
darkening due to the high density of defects induced by irradiation
(a photographic image of the implanted sample taken
before annealing is shown in Fig.~\ref{fig1}a).

In order to obtain optical activation of the SiV$^-$ centers,
the sample was annealed for 1 h at 1100$^\circ$ C in reducing atmosphere
(Ar + 2\% H$_{2}$). After annealing an almost complete recovery of the
transparency of the sample was obtained, due to high temperature
mobility of the defects, which favors both the recombination of the
Frenkel couples and the formation of the SiV$^-$ color centers.\cite{pezzagna11}
A fluorescence map acquired at 293 K is shown in Fig.~\ref{fig1}b.

The luminescence properties of the SiV\textsuperscript{-} centers were
studied by excitation with the 647 nm line of a Kr ion laser,
collecting the back-diffused luminescent signal in confocal configuration
through a micro-objective with NA=0.28
and analyzing it by means of a single monocromator 
equipped with a notch filter and with a 300 lines/mm grating followed by a
charge-coupled device (CCD) detector. 
The instrumental resolution of the system is  0.3 nm in the range 
from 650 to 820 nm. The transversal spatial resolution
is 2 $\mu$m and the depth of field is about 15 $\mu$m,
which makes possible collecting luminescence maps with
micrometric resolution.

The sample was placed in a water-cooled
high-vacuum chamber and heated by a helicoidal heater wound
around a copper holder to probe temperatures up to 861 K. 
The temperature of the sample was measured by
a thermocouple placed on the holder a few millimeters apart.
A fused silica window allows both excitation and observation of the sample.
In all the measurements, the power of the excitation beam was kept
constant to 36 $\pm$ 1 mW, estimated 
taking into account the losses for reflection
both on the fused silica window and on the diamond
surface.\footnote{Some percent of variation of the integrated signal was observed
for movements along the sample of the order of a few micrometers.
Consequently, since during the temperature ramp a slight displacement
due to the thermal expansion of the holder takes place, at each
change of temperature great attention was devoted to 
the maximization of the luminescent signal in the direction of the optical axis 
($z$) and to the tracking of the position of the measurement point in the 
$xy$ plane. Moreover, measurements were taken and averaged on a matrix 
of $5 \times 5$ points with a pitch of 2 \ensuremath{\mu}m, to 
make sure that the center of the original measurement point was
included in the measurement area.}
A series of data at room temperature were also obtained
at different power levels, in order to evaluate the saturation power.

Figure~\ref{fig1}c shows the spectra acquired at different temperatures
from 293 to 861 K. 
We observed a main peak centered between 738 and 751 nm,
which can be ascribed to the luminescent of the SiV$^-$ center,
and the diamond Raman signal.
The central frequency, FWHM and amplitude of the signal show modifications
as temperature changes. A significant broadening of the luminescence
curve occurs with the temperature increase,
so that the long-wavelength tail of the spectrum
exceeds the limit of the observation window, especially at high temperatures.
In order to take into account the tails,
the fluorescence peaks were fitted with a
pseudo-Voigt distribution added to a linear background.
The overall luminescent signal has been evaluated as the sum of the 
the area of the fitted pseudo-Voigt and of the phonon sideband contribution.
It has been found that beyond about 700 K the pseudo-Voigt distribution
fits quite well the spectrum in all the range (excluding the
Raman peak of diamond, see also the insets of Fig.~\ref{fig1}c). 

\begin{figure}[htbp]
\begin{center}
\includegraphics[width=8.5cm]{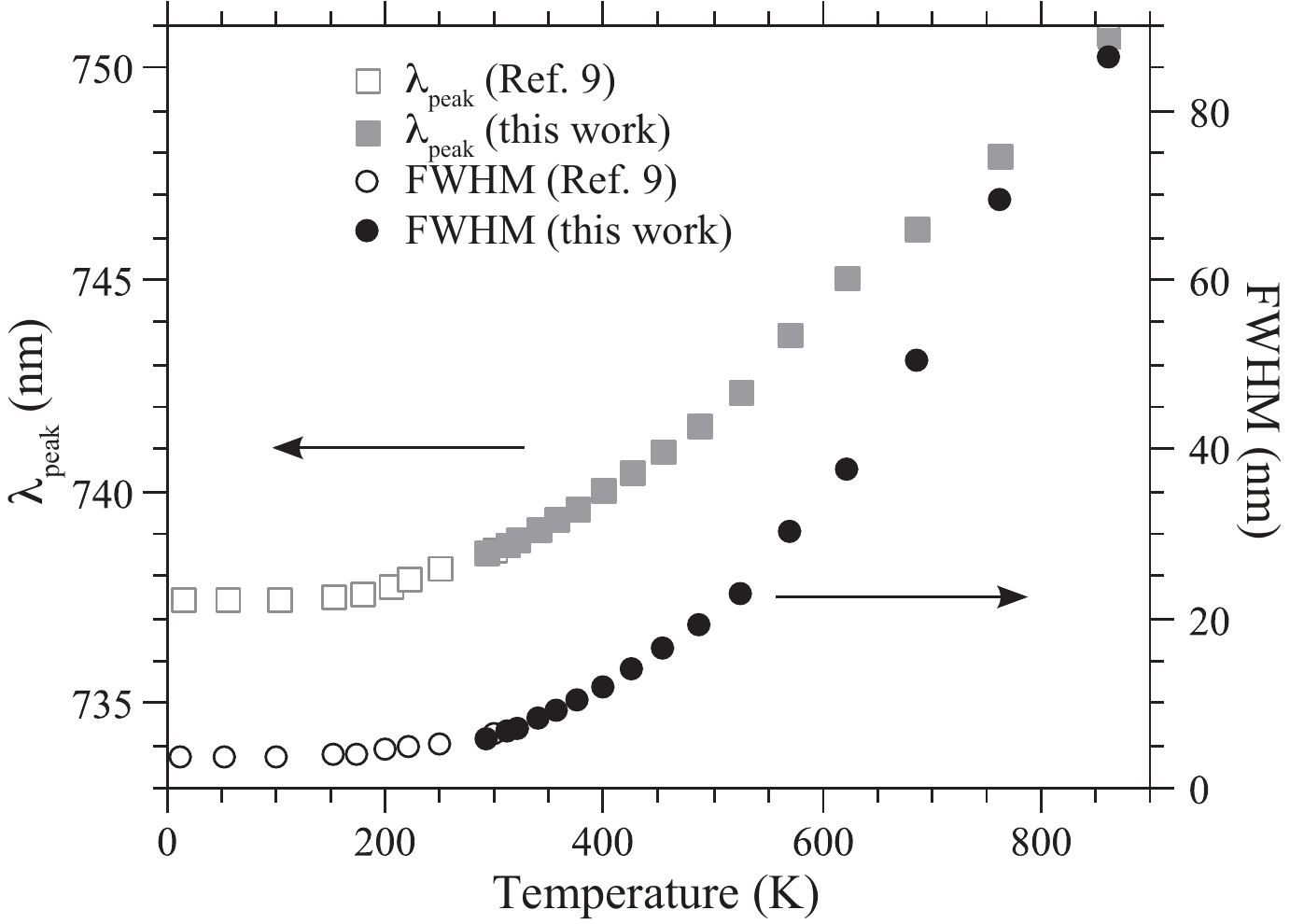}
\caption{\label{fig2}
Dependence of the central wavelength of the
luminescence peak $\lambda_\mathrm{peak}$ (squares, left vertical scale) and 
of the FWHM (circles, right vertical scale) on temperature.
Filled (open) symbols refer to measurements performed in
this work (in Ref.~\onlinecite{feng93}).}
\end{center}
\end{figure}

Figure~\ref{fig2} shows the temperature dependence of 
the central wavelength of the luminescence peak $\lambda_\mathrm{peak}$
and the FWHM of the fitted pseudo-Voigt distribution.
These are in excellent agreement with those observed by Feng and
Schwartz,\cite{feng93} confirmed also by
Neu et al.\cite{neu13} from cryogenic to room temperatures. 
The red shift, in particular, seems to be related to the shrink of the diamond 
band-gap,\cite{liu00}
which is likely to determine a change in the transition frequency 
of the deep levels inside the band-gap. 

\begin{figure}[htbp]
\begin{center}
\includegraphics[width=8.5cm]{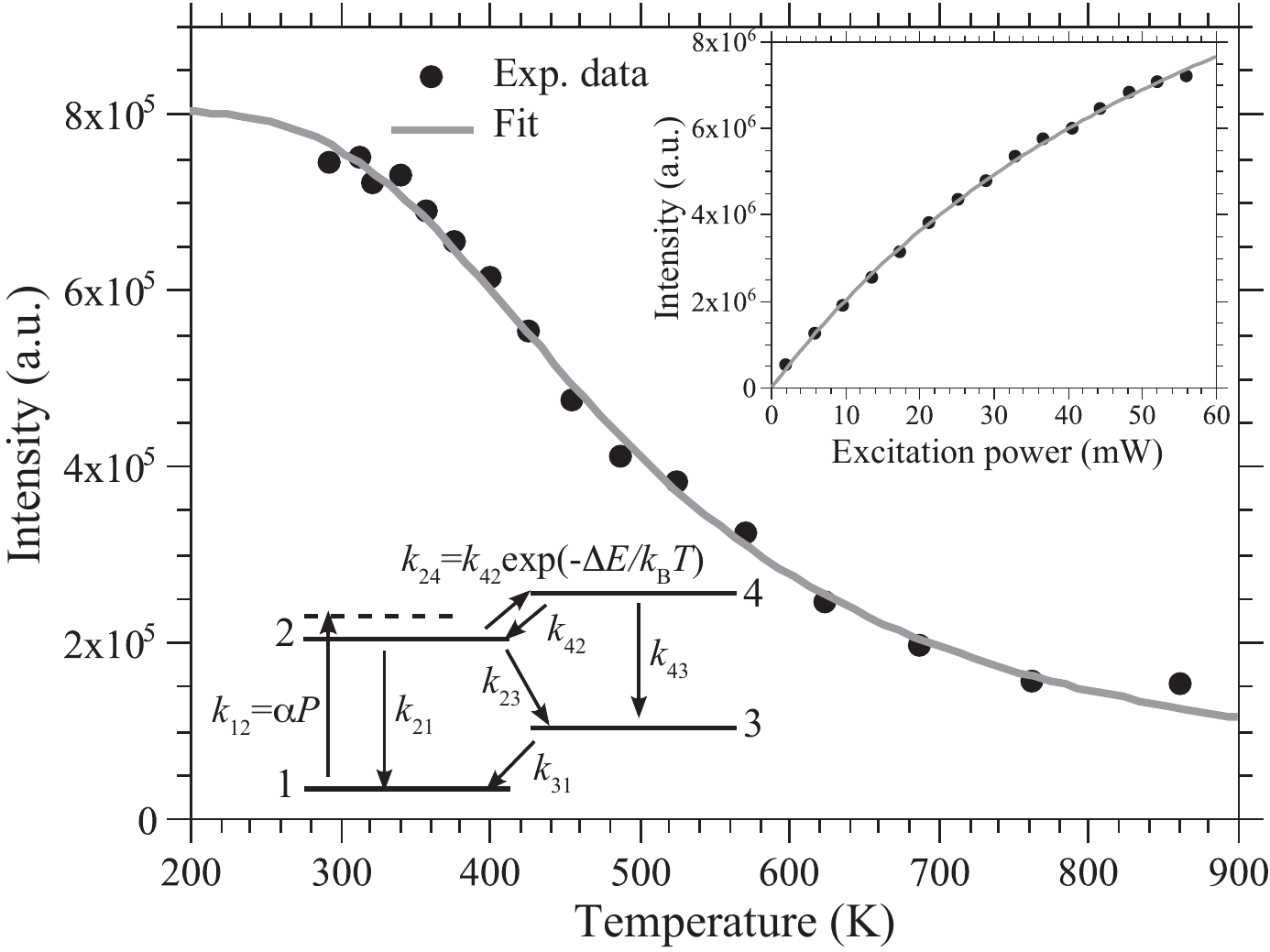}
\caption{\label{fig3}
Dots: experimental dependence on temperature of the integrated
luminescence of an ensemble of
SiV$^-$ centers. Line: fit of the experimental points by
means of Eq.~(\ref{eq1}), with $\Delta E = 180$ meV and $A_p=60$.
The inset in the upper-right part of Fig.~\ref{fig3}
shows the experimental dependence of the observed
luminescence on the excitation power at room temperature (dots) and the fit
of the data by means of the expression
$I = I_s \times P/(P_s +P)$, with $P_s$=75 mW (curve).
The bottom-left part shows the four-level scheme adopted to
explain the photophysics of the SiV$^-$ center at large temperatures.
We assumed a pump rate $k_{12}$ on the excitation level (dashed line),
which linearly depends on the laser power
(see Ref.~\onlinecite{neu12}) and $k_{24}$ bound to $k_{42}$ by the
Boltzmann factor for detailed balance considerations.}
\end{center}
\end{figure}

Figure~\ref{fig3} shows the behaviour of the integrated luminescence
signal with temperature (dots) along with the fit obtained with a
model described below. 
The quenching of the luminescence with increasing temperatures can be
explained assuming that a transition from the excited state occurs
with an activation energy $\Delta E$.
A large ensemble of different mechanisms leads to 
a same functional dependence on temperature $T$ of the fluorescence $I$,
so that the quenching at high temperatures can be expressed as\cite{feng93}
\begin{equation}
\label{eq1}
I=\frac{I_p}{1+A_p \exp(-\Delta E/k_BT)},
\end{equation}
where $I_p$ is the fluorescence at 0 K, $A_p$ is
a factor that depends on the model
and $k_B$ is the Boltzmann konstant.
Figure~\ref{fig3} compares the experimental dependence on
temperature of the integrated luminescence with Eq.~(\ref{eq1}),
showing very good agreement.



In order to better understand the photophysics
of the SiV$^-$ center, we explicitly take into account
saturation effects within a four-level scheme that will be
introduced later in more details (see the inset of Fig.~\ref{fig3}
and Appendix~\ref{4level}).
From the solution of the rate equations at equilibrium 
combined we obtain the following expressions for $I_p$ and $A_p$
in Eq.~(\ref{eq1})
\begin{equation}
\label{eq2}
I_p = I_s \frac{P}{P+P_s}; \hspace{1cm}   A_p=A\frac{P+P_0}{P+P_s}.
\end{equation}
Here, $A$ is a constant, $I_s$ is the saturation intensity 
at low temperatures, $P_s$ is the saturation 
power and $P_0$ is an onset excitation power corresponding to the 
quenching at low-excitation levels (see Appendix~\ref{4level} for details). 
All these parameters depend both on the 
rate constants of the system and on the coupling efficiency
between the individual center and the exciting radiation.

Experimental investigations performed on individual 
SiV$^-$ centers,\cite{neu12} in order to evaluate the characteristic 
rate constants of radiative and non-radiative transitions, show
that these are subject to considerable variations, presumably due to
different local environmental conditions.
Nevertheless, the comparison of Eq.~(\ref{eq1}) with our experimental data,
obtained from a large number of centers,
can lead to useful considerations about the average signal.

The inset in Fig.~\ref{fig3} shows the observed behavior of the integrated
luminescence as a function of the excitation power at 293 K.
The sublinear behavior of the luminescence with increasing 
power can be accounted for assuming
$I=I_s\times P/(P+P_s)$ at room temperature.
From the fit of the data it results that $P_s=75\pm 4$ mW.
Moreover, the fluorescence intensity versus temperature
is well fitted by Eq.~(\ref{eq1})
with $\Delta E$ and $A_p$ free parameters, which result to be
$\Delta E= 180\pm 10$ meV and $A_p=60\pm10$
(at the power $P=36$ mW). Such a high value of $A_p$
can serve as a discriminator between different alternative models of the
quenching mechanism.

For instance, if one assumes the simplest possible
mechanism based on three levels (see Appendix~\ref{3level}):
the ground (1) and the excited level (2)
at 1.68 eV, plus a third level (3) decaying to the
ground state and placed 0.18 eV above the excited 
level, 
the results are not compatible with the experimental numerical values.
Indeed, the form of the constant $A$ becomes
$A=k_{32}/(k_{32}+k_{31}) < 1$,
were $k_{ij}$ is the rate constant between the $i$ and the $j$ state. 
Here, the ratio $(P+P_0)/(P+P_s)$ would be comprised between 1 and 
$P_0/P_s=k_{31}/k_{21}$,
which is unlikely exceeding 1, given the strength of the radiative
transition $2 \to 1$. Thus, $A_p$ is hardly exceeding 1 in this scheme,
to be compared with $A_p = 60$ of our fit. 

On the other hand, a shelving 
state (3) placed \textit{below} the level 2 is assumed to explain the bounching 
behavior for large correlation times of SiV$^-$ centers as single photon
emitters,\cite{neu12} and a higher energy state radiatively bounded
to the shelving state is invoked to give account of the de-shelving
observed at high excitation powers.

It is then more conceivable a scheme with a shelving state (3) and a
fourth level (4), radiatively connected with the shelving state,
placed at 0.18 eV above the state (2),
as shown in the inset of Fig.~\ref{fig3},
being the transitions $2 \to 3$, $3 \to 1$ and $4 \to 2$ optically forbidden. 
The inversion symmetry of the D$_\mathrm{3d}$
point group implies indeed that the state functions of the SiV$^-$
have definite parity, giving a whole set of forbidden transitions of this
kind.\cite{gali13}
In this case, the constant $A$ assumes a quite complicated form,
which simplifies, in the case that the rates $k_{23}$ and $k_{31}$
are much smaller than $k_{21}$
(as in Ref.~\onlinecite{neu12}) in the following one:
$A=k_{43}/(k_{23}+k_{31})$.

Assuming that $k_{43} \sim k_{21}$ and referring to the average
values of $k_{21}$, $k_{23}$, $k_{31}$ reported in
Ref.~\onlinecite{neu12} (see table in the Appendix~\ref{rateconst}), we obtain $A \sim 170$.
Since the ratio $P_0/P_s=k_{23}/k_{21}$ is small 
compared with unity and given the fitted value of $P_s=75$ mW, the 
factor $(P+P_0)/(P+P_s)$ is approximately 0.32 and the 
product $A_p=A\times (P+P_0)/(P+P_s)$
results $\sim 55$, which has the correct order of magnitude. 

The possible emission corresponding to the transition $4 \to 3$ is governed
by an equation of the kind
\begin{equation}
I=I_\mathrm{ZPL}\frac{k_{42}k_{43}}{k_{21}(k_{42}+k_{43})}
\exp(-\Delta E/k_BT),
\end{equation}
which gives a negligible signal at room temperature 
and also at 800 K does not exceed 1\% of the $I_\mathrm{ZPL}$
at room temperature.

In conclusion, we discovered that the integrated luminescence 
of the SiV$^-$ color center is still strong at temperatures as high as 500 K,
with a FWHM which remains lower than 20 nm.
We attributed such a relatively weak decrease of the fluorescence signal
with increasing temperature to population transfer to a shelving state,
enhanced by a transition channel with an activation energy of 180 meV.
Our findings provide distinct information on the optical
properties of the SiV$^-$, which may be relevant for developing
light-emitting devices that operate at high temperatures, 
and encourage further investigations of color centers in diamond 
under such experimental conditions.

%
%

%


Financial support from the EC Seventh Framework Programme (248855),
the Helmoltz Association (283286) and the Ministry of Education and
Science of the Russian Federation (16.19.2014/K) is gratefully acknowledged.
The authors would like to thank
K. De Hanstetters and S. Orlanducci for helpful conversations and
for providing diamond samples, 
L. Ulivi for the use of the annealing chamber.

\appendix

\section{Four-level scheme}
\label{4level}

The model referred to in the main text is described by the following four-level scheme
\begin{figure}[htbp]
\begin{center}
\includegraphics[width=5cm]{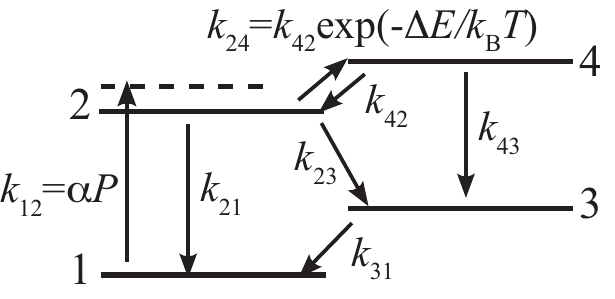}
\caption{\label{fig1-supp} Four-level scheme that explains the experimental obeservations.}
\end{center}
\end{figure}

The rate equations, at equilibrium, read
\begin{eqnarray}
k_{12}n_1-\left(k_{21}+k_{24}+k_{23}\right)n_2+k_{42}n_4 & = & 0, \\
k_{23}n_2-k_{31}n_3+k_{43}n_4 & = & 0, \\
k_{24}n_2-\left(k_{42}+k_{43}\right)n_4 & = & 0, \\
n_1+n_2+n_3+n_4 & = & 1.
\end{eqnarray}

The first three equations describe the stationary populations 
of the levels 2, 3 and 4, the last one normalizes the sum of the occupation probability 
of the four levels.

We assume $k_{12}=\alpha P$, because 
the vibrationally excited states populated by the exciting radiation are kept weakly 
populated by fast thermal de-excitation to the level 2, avoiding stimulated emission 
and saturation on the pumping transition. The ratio of the two thermal transition 
rates $k_{24}$ and $k_{42}$ is 
fixed to the Boltzmann factor by detailed balance consideration, giving
$k_{24}=k_{42}\exp\left(-\Delta E/k_B T\right)$
where $\Delta E$ is the activation energy of the transition and
$k_B$ is the Boltzmann constant.

Solving the system of rate equations we found that the occupation of the level 
2, proportional to the luminescence intensity, is given by
\begin{equation}
\label{n2}
n_2=n_s\frac{P}{P+P_s}\frac{1}{1+A\frac{P+P_0}{P+P_s}\exp
\left(-\frac{\Delta E}{k_B T}\right)},
\end{equation}
which corresponds to Eq.~(1) of the main text.
The constants $n_s$, $P_0$, $P_s$ e $A$ are bounded to the rate constants 
by the following relations
\begin{eqnarray}
n_s & = & \frac{k_{31}}{k_{31}+k_{23}}, \\
P_0 & = & \frac{k_{31}k_{43}}{\alpha\left(k_{31}+k_{43}\right)}, \\
P_s & = & \frac{k_{31}\left(k_{21}+k_{23}\right)}{\alpha\left(k_{31}+k_{23}\right)}, \\
A & = & \frac{k_{42}\left(k_{31}+k_{43}\right)}{\left(k_{42}+k_{43}\right)
\left(k_{31}+k_{23}\right)}.
\end{eqnarray}

The transition energy $\Delta E$ ($180\pm 10$ meV) has the same order of magnitude
of the diamond optical phonon energy (165 meV),\cite{schwoerer98} which makes plausible the
inequality $k_{42} \gg k_{43}$. The transition $3\to 1$, on the other hand, is optically forbidden,
thus $k_{43} \gg k_{31}$. As a consequence $A\simeq k_{43}/\left(k_{31}+k_{23}\right)$,
which is the approximation referred to in the main text.

Another constant referred to in the main text is the ratio
\begin{equation}
\frac{P_0}{P_s}=\frac{k_{43}\left(k_{31}+k_{23}\right)}{\left(k_{31}+k_{43}\right)
\left(k_{21}+k_{23}\right)}.
\end{equation}

Under the above mentioned hypothesis, $k_{43} \gg k_{31}$, and considering
$k_{21} \gg k_{23} \gg k_{31}$, which results from the data reported in Ref.~\onlinecite{neu12}
and reproduced in Tab.~\ref{tab1}, this ratio reduces to $P_0/P_s\simeq
k_{23}/k_{21}$, given in the main text.

If the transition $4\to 3$ is optical, which could be the case 
if the levels 1 and 3 had opposed parity with respect to the
levels 2 and 4, one can wonder if a thermally enhanced 
luminescent signal from transition $4\to 3$ could be detected. This is probably not 
the case, inasmuch the population of the level 4 (according to the third rate equation)
is proportional to $n_2$ according to the constant $k_{24}/\left(k_{42}+k_{43}\right)$.
Consequently, considering the signal $I_\mathrm{ZPL}$
at the zero-phonon line (ZPL) of the 738 nm transition
to be proportional to $k_{21}n_2$ 
and the hypothetical signal $I_\mathrm{hyp}$ from the transition $4\to 3$ 
to be proportional to $k_{43}n_4$, 
one finds that the two signals should be in the ratio
\begin{equation}
\frac{I_\mathrm{hyp}}{I_\mathrm{ZPL}}=
\frac{k_{42}k_{43}}{k_{21}\left(k_{42}+k_{43}\right)}
\exp\left(-\frac{\Delta E}{k_B T}\right).
\end{equation}

The Boltzman factor in the previous equation is negligible at 
room temperature and at the highest temperatures experimented, when it reaches the 
value 0.085, the 738 nm luminescence is so low to make unlikely the detection of 
a signal more than 10 times lower.

One can extend the calculations above to the case in which the 
transition $4\to 1$ is permitted (as assumed, for example, in Ref.~\onlinecite{neu12})
adding a finite transition rate $k_{41}$ and
writing the third rate equation of the system as
\begin{equation}
k_{24}n_2-\left(k_{41}+k_{42}+k_{43}\right)n_4=0.
\end{equation}

In this case, the $A$ constant is written in the following way:
\begin{equation}
A=\frac{k_{42}\left(k_{31}+k_{43}\right)}{\left(k_{41}+k_{42}+k_{43}\right)
\left(k_{31}+k_{23}\right)}.
\end{equation}

If the thermal transition rate $k_{42}$ is much greater not only
than $k_{43}$ (as assumed earlier) but also of $k_{41}$, 
we will have $A\simeq \frac{k_{43}}{\left(k_{31}+k_{23}\right)}$ as in the previous case.
The ratio $P_0/P_s$ on the other hand, is now given by
\begin{equation}
\frac{P_0}{P_s}=\frac{\left(k_{43}+k_{41}\right)\left(k_{31}+k_{23}\right)}
{\left(k_{31}+k_{43}\right)\left(k_{21}+k_{23}\right)},
\end{equation}
giving again $P_0/P_s\simeq k_{23}/k_{21}$.

\newpage
\subsection{Experimental rate constants}
\label{rateconst}

\begin{table}[h!]
\begin{center}
\begin{tabular}{|c|c|c|c|}
\hline
$k_{21}$ & $k_{23}$ & $k_{31}$ & $k_{21}/\left(k_{23}+k_{31}\right)$  \\
\hline
4408 & 137 & 0.27 & 32.1 \\
\hline
3424 & 24.6 & 1.7 & 130.2 \\
\hline
771 & 23.6 & 0.35 & 32.2 \\
\hline
1084 & 31.7 & 0.12 & 34.1 \\
\hline
1545.1 & 17.4 & 1 & 84.0 \\
\hline
770.1 & 11.1 & 0.81 & 64.7 \\
\hline
1053.6 & 21.7 & 0.13 & 48.3 \\
\hline
3479 & 92.6 & 0.82 & 37.2 \\
\hline
161 & 7.3 & 0.24 & 21.4 \\
\hline
1638 & 1.5 & 0.16 & 986.7 \\
\hline
2487 & 12.5 & 0.15 & 196.6 \\
\hline
1181.7 & 1.8 & 0.23 & 582.1 \\
\hline
798.8 & 34.6 & 0.24 & 22.9 \\
\hline
1076 & 13.3 & 0.32 & 79.0 \\
\hline
\end{tabular}
\caption{\label{tab1} Values of the constant $k_{ij}$ measured in Ref.~\onlinecite{neu12}
for a sample of 14 SiV$^-$ centers in nanodiamond (the first 7 rows of the table)
and nanoislands of diamond grown on iridium (the following 7 rows).
The average value of the ratio $k_{21}/\left(k_{23}+k_{23}\right)$ is 
$170\pm 75$, which, assuming $k_{43}\sim k_{21}$, should give at least the correct
order of $A=k_{43}/\left(k_{23}+k_{23}\right)$.}
\end{center}
\end{table}

\section{Three-level scheme}
\label{3level}

\begin{figure}[htbp]
\begin{center}
\includegraphics[width=5cm]{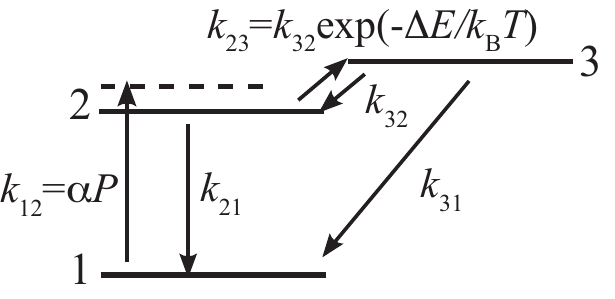}
\caption{\label{fig2-supp} Three-level scheme discussed in the main text.}
\end{center}
\end{figure}

A much simpler model referred to in the main text is given by the two levels 1 and 
2 plus a third one decaying to the ground state and populated by 
thermal stimulation from the level 2 (see Fig.~\ref{fig2-supp}). 

The solution of the rate equations gives a dependence of the population $n_2$ on
temperature $T$ and power $P$ of the same kind described by Eq.~(\ref{n2}),
but with the following parameters $n_s=1$; $P_0=k_{31}/\alpha$;
$P_s=k_{21}/\alpha$ and $A=k_{32}/\left(k_{31}+k_{32}\right)$.
As already mentioned in the main text, the value of $A$, which cannot exceed
unity, is hardly compatible with the values suggested by the fit of the experimental data.


\begin{thebibliography}{18}%
\makeatletter
\providecommand \@ifxundefined [1]{%
 \@ifx{#1\undefined}
}%
\providecommand \@ifnum [1]{%
 \ifnum #1\expandafter \@firstoftwo
 \else \expandafter \@secondoftwo
 \fi
}%
\providecommand \@ifx [1]{%
 \ifx #1\expandafter \@firstoftwo
 \else \expandafter \@secondoftwo
 \fi
}%
\providecommand \natexlab [1]{#1}%
\providecommand \enquote  [1]{``#1''}%
\providecommand \bibnamefont  [1]{#1}%
\providecommand \bibfnamefont [1]{#1}%
\providecommand \citenamefont [1]{#1}%
\providecommand \href@noop [0]{\@secondoftwo}%
\providecommand \href [0]{\begingroup \@sanitize@url \@href}%
\providecommand \@href[1]{\@@startlink{#1}\@@href}%
\providecommand \@@href[1]{\endgroup#1\@@endlink}%
\providecommand \@sanitize@url [0]{\catcode `\\12\catcode `\$12\catcode
  `\&12\catcode `\#12\catcode `\^12\catcode `\_12\catcode `\%12\relax}%
\providecommand \@@startlink[1]{}%
\providecommand \@@endlink[0]{}%
\providecommand \url  [0]{\begingroup\@sanitize@url \@url }%
\providecommand \@url [1]{\endgroup\@href {#1}{\urlprefix }}%
\providecommand \urlprefix  [0]{URL }%
\providecommand \Eprint [0]{\href }%
\providecommand \doibase [0]{http://dx.doi.org/}%
\providecommand \selectlanguage [0]{\@gobble}%
\providecommand \bibinfo  [0]{\@secondoftwo}%
\providecommand \bibfield  [0]{\@secondoftwo}%
\providecommand \translation [1]{[#1]}%
\providecommand \BibitemOpen [0]{}%
\providecommand \bibitemStop [0]{}%
\providecommand \bibitemNoStop [0]{.\EOS\space}%
\providecommand \EOS [0]{\spacefactor3000\relax}%
\providecommand \BibitemShut  [1]{\csname bibitem#1\endcsname}%
\let\auto@bib@innerbib\@empty
\bibitem [{\citenamefont {Willander}\ \emph {et~al.}(2006)\citenamefont
  {Willander}, \citenamefont {Friesel}, \citenamefont {Wahab},\ and\
  \citenamefont {Straumal}}]{willander06}%
  \BibitemOpen
  \bibfield  {author} {\bibinfo {author} {\bibfnamefont {M.}~\bibnamefont
  {Willander}}, \bibinfo {author} {\bibfnamefont {M.}~\bibnamefont {Friesel}},
  \bibinfo {author} {\bibfnamefont {Q.-U.}\ \bibnamefont {Wahab}}, \ and\
  \bibinfo {author} {\bibfnamefont {B.}~\bibnamefont {Straumal}},\ }\href
  {\doibase 10.1007/s10854-005-5137-4} {\bibfield  {journal} {\bibinfo
  {journal} {J. Mater. Science: Materials in Electronics}\ }\textbf {\bibinfo
  {volume} {17}},\ \bibinfo {pages} {1} (\bibinfo {year} {2006})}\BibitemShut
  {NoStop}%
\bibitem [{\citenamefont {Lohrmann}\ \emph {et~al.}(2011)\citenamefont
  {Lohrmann}, \citenamefont {Pezzagna}, \citenamefont {Dobrinets},
  \citenamefont {Spinicelli}, \citenamefont {Jacques}, \citenamefont {Roch},
  \citenamefont {Meijer},\ and\ \citenamefont {Zaitsev}}]{lohrmann11}%
  \BibitemOpen
  \bibfield  {author} {\bibinfo {author} {\bibfnamefont {A.}~\bibnamefont
  {Lohrmann}}, \bibinfo {author} {\bibfnamefont {S.}~\bibnamefont {Pezzagna}},
  \bibinfo {author} {\bibfnamefont {I.}~\bibnamefont {Dobrinets}}, \bibinfo
  {author} {\bibfnamefont {P.}~\bibnamefont {Spinicelli}}, \bibinfo {author}
  {\bibfnamefont {V.}~\bibnamefont {Jacques}}, \bibinfo {author} {\bibfnamefont
  {J.-F.}\ \bibnamefont {Roch}}, \bibinfo {author} {\bibfnamefont
  {J.}~\bibnamefont {Meijer}}, \ and\ \bibinfo {author} {\bibfnamefont {A.~M.}\
  \bibnamefont {Zaitsev}},\ }\href {\doibase
  http://dx.doi.org/10.1063/1.3670332} {\bibfield  {journal} {\bibinfo
  {journal} {Appl. Phys. Lett.}\ }\textbf {\bibinfo {volume} {99}},\ \bibinfo
  {eid} {251106} (\bibinfo {year} {2011})}\BibitemShut {NoStop}%
\bibitem [{\citenamefont {Mizuochi}\ \emph {et~al.}(2012)\citenamefont
  {Mizuochi}, \citenamefont {Makino}, \citenamefont {Kato}, \citenamefont
  {Takeuchi}, \citenamefont {Ogura}, \citenamefont {Okushi}, \citenamefont
  {Nothaft}, \citenamefont {Neumann}, \citenamefont {Gali}, \citenamefont
  {Jelezko}, \citenamefont {Wrachtrup},\ and\ \citenamefont
  {Yamasaki}}]{mizuoki12}%
  \BibitemOpen
  \bibfield  {author} {\bibinfo {author} {\bibfnamefont {N.}~\bibnamefont
  {Mizuochi}}, \bibinfo {author} {\bibfnamefont {T.}~\bibnamefont {Makino}},
  \bibinfo {author} {\bibfnamefont {H.}~\bibnamefont {Kato}}, \bibinfo {author}
  {\bibfnamefont {D.}~\bibnamefont {Takeuchi}}, \bibinfo {author}
  {\bibfnamefont {M.}~\bibnamefont {Ogura}}, \bibinfo {author} {\bibfnamefont
  {H.}~\bibnamefont {Okushi}}, \bibinfo {author} {\bibfnamefont
  {M.}~\bibnamefont {Nothaft}}, \bibinfo {author} {\bibfnamefont
  {P.}~\bibnamefont {Neumann}}, \bibinfo {author} {\bibfnamefont
  {A.}~\bibnamefont {Gali}}, \bibinfo {author} {\bibfnamefont {F.}~\bibnamefont
  {Jelezko}}, \bibinfo {author} {\bibfnamefont {J.}~\bibnamefont {Wrachtrup}},
  \ and\ \bibinfo {author} {\bibfnamefont {S.}~\bibnamefont {Yamasaki}},\
  }\href {\doibase http://dx.doi.org/10.1038/nphoton.2012.75} {\bibfield
  {journal} {\bibinfo  {journal} {Nat. Photon.}\ }\textbf {\bibinfo {volume}
  {6}},\ \bibinfo {pages} {299} (\bibinfo {year} {2012})}\BibitemShut {NoStop}%
\bibitem [{\citenamefont {Berhane}\ \emph {et~al.}(2015)\citenamefont
  {Berhane}, \citenamefont {Choi}, \citenamefont {Kato}, \citenamefont
  {Makino}, \citenamefont {Mizuochi}, \citenamefont {Yamasaki},\ and\
  \citenamefont {Aharonovich}}]{berhane15}%
  \BibitemOpen
  \bibfield  {author} {\bibinfo {author} {\bibfnamefont {A.~M.}\ \bibnamefont
  {Berhane}}, \bibinfo {author} {\bibfnamefont {S.}~\bibnamefont {Choi}},
  \bibinfo {author} {\bibfnamefont {H.}~\bibnamefont {Kato}}, \bibinfo {author}
  {\bibfnamefont {T.}~\bibnamefont {Makino}}, \bibinfo {author} {\bibfnamefont
  {N.}~\bibnamefont {Mizuochi}}, \bibinfo {author} {\bibfnamefont
  {S.}~\bibnamefont {Yamasaki}}, \ and\ \bibinfo {author} {\bibfnamefont
  {I.}~\bibnamefont {Aharonovich}},\ }\href {\doibase
  http://dx.doi.org/10.1063/1.4919388} {\bibfield  {journal} {\bibinfo
  {journal} {Appl. Phys. Lett.}\ }\textbf {\bibinfo {volume} {106}},\ \bibinfo
  {eid} {171102} (\bibinfo {year} {2015})}\BibitemShut {NoStop}%
\bibitem [{\citenamefont {Lohrmann}\ \emph {et~al.}(2015)\citenamefont
  {Lohrmann}, \citenamefont {Iwamoto}, \citenamefont {Bodrog}, \citenamefont
  {Castelletto}, \citenamefont {Ohshima}, \citenamefont {Karle}, \citenamefont
  {Gali}, \citenamefont {Prawer}, \citenamefont {McCallum},\ and\ \citenamefont
  {Johnson}}]{lohrmann15}%
  \BibitemOpen
  \bibfield  {author} {\bibinfo {author} {\bibfnamefont {A.}~\bibnamefont
  {Lohrmann}}, \bibinfo {author} {\bibfnamefont {N.}~\bibnamefont {Iwamoto}},
  \bibinfo {author} {\bibfnamefont {Z.}~\bibnamefont {Bodrog}}, \bibinfo
  {author} {\bibfnamefont {S.}~\bibnamefont {Castelletto}}, \bibinfo {author}
  {\bibfnamefont {T.}~\bibnamefont {Ohshima}}, \bibinfo {author} {\bibfnamefont
  {T.}~\bibnamefont {Karle}}, \bibinfo {author} {\bibfnamefont
  {A.}~\bibnamefont {Gali}}, \bibinfo {author} {\bibfnamefont {S.}~\bibnamefont
  {Prawer}}, \bibinfo {author} {\bibfnamefont {J.}~\bibnamefont {McCallum}}, \
  and\ \bibinfo {author} {\bibfnamefont {B.}~\bibnamefont {Johnson}},\
  }\href@noop {} {\bibfield  {journal} {\bibinfo  {journal} {Nat. Commun.}\
  }\textbf {\bibinfo {volume} {6}},\ \bibinfo {pages} {7783} (\bibinfo {year}
  {2015})}\BibitemShut {NoStop}%
\bibitem [{\citenamefont {Gabrysch}\ \emph {et~al.}(2008)\citenamefont
  {Gabrysch}, \citenamefont {Majdi}, \citenamefont {Hall\'{e}n}, \citenamefont
  {Linnarsson}, \citenamefont {Sch\"{o}ner}, \citenamefont {Twitchen},\ and\
  \citenamefont {Isberg}}]{gabrysch08}%
  \BibitemOpen
  \bibfield  {author} {\bibinfo {author} {\bibfnamefont {M.}~\bibnamefont
  {Gabrysch}}, \bibinfo {author} {\bibfnamefont {S.}~\bibnamefont {Majdi}},
  \bibinfo {author} {\bibfnamefont {A.}~\bibnamefont {Hall\'{e}n}}, \bibinfo
  {author} {\bibfnamefont {M.}~\bibnamefont {Linnarsson}}, \bibinfo {author}
  {\bibfnamefont {A.}~\bibnamefont {Sch\"{o}ner}}, \bibinfo {author}
  {\bibfnamefont {D.}~\bibnamefont {Twitchen}}, \ and\ \bibinfo {author}
  {\bibfnamefont {J.}~\bibnamefont {Isberg}},\ }\href {\doibase
  10.1002/pssa.200879711} {\bibfield  {journal} {\bibinfo  {journal} {Phys.
  Stat. Sol. (a)}\ }\textbf {\bibinfo {volume} {205}},\ \bibinfo {pages} {2190}
  (\bibinfo {year} {2008})}\BibitemShut {NoStop}%
\bibitem [{\citenamefont {Stenger}\ \emph {et~al.}(2013)\citenamefont
  {Stenger}, \citenamefont {Pinault-Thaury}, \citenamefont {Kociniewski},
  \citenamefont {Lusson}, \citenamefont {Chikoidze}, \citenamefont {Jomard},
  \citenamefont {Dumont}, \citenamefont {Chevallier},\ and\ \citenamefont
  {Barjon}}]{stenger13}%
  \BibitemOpen
  \bibfield  {author} {\bibinfo {author} {\bibfnamefont {I.}~\bibnamefont
  {Stenger}}, \bibinfo {author} {\bibfnamefont {M.-A.}\ \bibnamefont
  {Pinault-Thaury}}, \bibinfo {author} {\bibfnamefont {T.}~\bibnamefont
  {Kociniewski}}, \bibinfo {author} {\bibfnamefont {A.}~\bibnamefont {Lusson}},
  \bibinfo {author} {\bibfnamefont {E.}~\bibnamefont {Chikoidze}}, \bibinfo
  {author} {\bibfnamefont {F.}~\bibnamefont {Jomard}}, \bibinfo {author}
  {\bibfnamefont {Y.}~\bibnamefont {Dumont}}, \bibinfo {author} {\bibfnamefont
  {J.}~\bibnamefont {Chevallier}}, \ and\ \bibinfo {author} {\bibfnamefont
  {J.}~\bibnamefont {Barjon}},\ }\href {\doibase
  http://dx.doi.org/10.1063/1.4818946} {\bibfield  {journal} {\bibinfo
  {journal} {J. Appl. Phys.}\ }\textbf {\bibinfo {volume} {114}},\ \bibinfo
  {eid} {073711} (\bibinfo {year} {2013})}\BibitemShut {NoStop}%
\bibitem [{\citenamefont {Pezzagna}\ \emph {et~al.}(2011)\citenamefont
  {Pezzagna}, \citenamefont {Rogalla}, \citenamefont {Wildanger}, \citenamefont
  {Meijer},\ and\ \citenamefont {Zaitsev}}]{pezzagna11}%
  \BibitemOpen
  \bibfield  {author} {\bibinfo {author} {\bibfnamefont {S.}~\bibnamefont
  {Pezzagna}}, \bibinfo {author} {\bibfnamefont {D.}~\bibnamefont {Rogalla}},
  \bibinfo {author} {\bibfnamefont {D.}~\bibnamefont {Wildanger}}, \bibinfo
  {author} {\bibfnamefont {J.}~\bibnamefont {Meijer}}, \ and\ \bibinfo {author}
  {\bibfnamefont {A.}~\bibnamefont {Zaitsev}},\ }\href
  {http://stacks.iop.org/1367-2630/13/i=3/a=035024} {\bibfield  {journal}
  {\bibinfo  {journal} {New J. Phys.}\ }\textbf {\bibinfo {volume} {13}},\
  \bibinfo {pages} {035024} (\bibinfo {year} {2011})}\BibitemShut {NoStop}%
\bibitem [{\citenamefont {Feng}\ and\ \citenamefont {Schwartz}(1993)}]{feng93}%
  \BibitemOpen
  \bibfield  {author} {\bibinfo {author} {\bibfnamefont {T.}~\bibnamefont
  {Feng}}\ and\ \bibinfo {author} {\bibfnamefont {B.~D.}\ \bibnamefont
  {Schwartz}},\ }\href {\doibase http://dx.doi.org/10.1063/1.353239} {\bibfield
   {journal} {\bibinfo  {journal} {J. Appl. Phys.}\ }\textbf {\bibinfo {volume}
  {73}},\ \bibinfo {pages} {1415} (\bibinfo {year} {1993})}\BibitemShut
  {NoStop}%
\bibitem [{\citenamefont {Gorokhovsky}\ \emph {et~al.}(1995)\citenamefont
  {Gorokhovsky}, \citenamefont {Turukhin}, \citenamefont {Alfano},\ and\
  \citenamefont {Phillips}}]{gorokhovsky95}%
  \BibitemOpen
  \bibfield  {author} {\bibinfo {author} {\bibfnamefont {A.~A.}\ \bibnamefont
  {Gorokhovsky}}, \bibinfo {author} {\bibfnamefont {A.~V.}\ \bibnamefont
  {Turukhin}}, \bibinfo {author} {\bibfnamefont {R.~R.}\ \bibnamefont
  {Alfano}}, \ and\ \bibinfo {author} {\bibfnamefont {W.}~\bibnamefont
  {Phillips}},\ }\href {\doibase http://dx.doi.org/10.1063/1.114176} {\bibfield
   {journal} {\bibinfo  {journal} {Appl. Phys. Lett.}\ }\textbf {\bibinfo
  {volume} {66}},\ \bibinfo {pages} {43} (\bibinfo {year} {1995})}\BibitemShut
  {NoStop}%
\bibitem [{\citenamefont {Neu}\ \emph {et~al.}(2013)\citenamefont {Neu},
  \citenamefont {Hepp}, \citenamefont {Hauschild}, \citenamefont {Gsell},
  \citenamefont {Fischer}, \citenamefont {Sternschulte}, \citenamefont
  {Steinm\"uller-Nethl}, \citenamefont {Schreck},\ and\ \citenamefont
  {Becher}}]{neu13}%
  \BibitemOpen
  \bibfield  {author} {\bibinfo {author} {\bibfnamefont {E.}~\bibnamefont
  {Neu}}, \bibinfo {author} {\bibfnamefont {C.}~\bibnamefont {Hepp}}, \bibinfo
  {author} {\bibfnamefont {M.}~\bibnamefont {Hauschild}}, \bibinfo {author}
  {\bibfnamefont {S.}~\bibnamefont {Gsell}}, \bibinfo {author} {\bibfnamefont
  {M.}~\bibnamefont {Fischer}}, \bibinfo {author} {\bibfnamefont
  {H.}~\bibnamefont {Sternschulte}}, \bibinfo {author} {\bibfnamefont
  {D.}~\bibnamefont {Steinm\"uller-Nethl}}, \bibinfo {author} {\bibfnamefont
  {M.}~\bibnamefont {Schreck}}, \ and\ \bibinfo {author} {\bibfnamefont
  {C.}~\bibnamefont {Becher}},\ }\href
  {http://stacks.iop.org/1367-2630/15/i=4/a=043005} {\bibfield  {journal}
  {\bibinfo  {journal} {New J. Phys.}\ }\textbf {\bibinfo {volume} {15}},\
  \bibinfo {pages} {043005} (\bibinfo {year} {2013})}\BibitemShut {NoStop}%
\bibitem [{\citenamefont {Neu}, \citenamefont {Agio},\ and\ \citenamefont
  {Becher}(2012)}]{neu12}%
  \BibitemOpen
  \bibfield  {author} {\bibinfo {author} {\bibfnamefont {E.}~\bibnamefont
  {Neu}}, \bibinfo {author} {\bibfnamefont {M.}~\bibnamefont {Agio}}, \ and\
  \bibinfo {author} {\bibfnamefont {C.}~\bibnamefont {Becher}},\ }\href
  {\doibase 10.1364/OE.20.019956} {\bibfield  {journal} {\bibinfo  {journal}
  {Opt. Express}\ }\textbf {\bibinfo {volume} {20}},\ \bibinfo {pages} {19956}
  (\bibinfo {year} {2012})}\BibitemShut {NoStop}%
\bibitem [{\citenamefont {Palla}\ \emph {et~al.}(2015)\citenamefont {Palla},
  \citenamefont {Czelusniak}, \citenamefont {Taccetti}, \citenamefont
  {Carraresi}, \citenamefont {Castelli}, \citenamefont {Fedi}, \citenamefont
  {Giuntini}, \citenamefont {Maurenzig}, \citenamefont {Sottili},\ and\
  \citenamefont {Taccetti}}]{palla15}%
  \BibitemOpen
  \bibfield  {author} {\bibinfo {author} {\bibfnamefont {L.}~\bibnamefont
  {Palla}}, \bibinfo {author} {\bibfnamefont {C.}~\bibnamefont {Czelusniak}},
  \bibinfo {author} {\bibfnamefont {F.}~\bibnamefont {Taccetti}}, \bibinfo
  {author} {\bibfnamefont {L.}~\bibnamefont {Carraresi}}, \bibinfo {author}
  {\bibfnamefont {L.}~\bibnamefont {Castelli}}, \bibinfo {author}
  {\bibfnamefont {M.}~\bibnamefont {Fedi}}, \bibinfo {author} {\bibfnamefont
  {L.}~\bibnamefont {Giuntini}}, \bibinfo {author} {\bibfnamefont
  {P.}~\bibnamefont {Maurenzig}}, \bibinfo {author} {\bibfnamefont
  {L.}~\bibnamefont {Sottili}}, \ and\ \bibinfo {author} {\bibfnamefont
  {N.}~\bibnamefont {Taccetti}},\ }\href {\doibase 10.1140/epjp/i2015-15039-y}
  {\bibfield  {journal} {\bibinfo  {journal} {European Phys. J. Plus}\ }\textbf
  {\bibinfo {volume} {130}},\ \bibinfo {eid} {39} (\bibinfo {year} {2015}),\
  10.1140/epjp/i2015-15039-y}\BibitemShut {NoStop}%
\bibitem [{\citenamefont {Ziegler}, \citenamefont {Biersack},\ and\
  \citenamefont {Littmark}(1985)}]{ziegler85}%
  \BibitemOpen
  \bibfield  {author} {\bibinfo {author} {\bibfnamefont {J.}~\bibnamefont
  {Ziegler}}, \bibinfo {author} {\bibfnamefont {J.}~\bibnamefont {Biersack}}, \
  and\ \bibinfo {author} {\bibfnamefont {U.}~\bibnamefont {Littmark}},\
  }\href@noop {} {\emph {\bibinfo {title} {The Stopping and Range of Ions in
  Solids}}},\ Stopping and ranges of ions in matter\ (\bibinfo  {publisher}
  {Pergamon},\ \bibinfo {year} {1985})\BibitemShut {NoStop}%
\bibitem [{Note1()}]{Note1}%
  \BibitemOpen
  \bibinfo {note} {Some percent of variation of the integrated signal was
  observed for movements along the sample of the order of a few micrometers.
  Consequently, since during the temperature ramp a slight displacement due to
  the thermal expansion of the holder takes place, at each change of
  temperature great attention was devoted to the maximization of the
  luminescent signal in the direction of the optical axis ($z$) and to the
  tracking of the position of the measurement point in the $xy$ plane.
  Moreover, measurements were taken and averaged on a matrix of $5 \times 5$
  points with a pitch of 2 \protect \ensuremath {\mu }m, to make sure that the
  center of the original measurement point was included in the measurement
  area.}\BibitemShut {Stop}%
\bibitem [{\citenamefont {Liu}\ \emph {et~al.}(2000)\citenamefont {Liu},
  \citenamefont {Bursill}, \citenamefont {Prawer},\ and\ \citenamefont
  {Beserman}}]{liu00}%
  \BibitemOpen
  \bibfield  {author} {\bibinfo {author} {\bibfnamefont {M.~S.}\ \bibnamefont
  {Liu}}, \bibinfo {author} {\bibfnamefont {L.~A.}\ \bibnamefont {Bursill}},
  \bibinfo {author} {\bibfnamefont {S.}~\bibnamefont {Prawer}}, \ and\ \bibinfo
  {author} {\bibfnamefont {R.}~\bibnamefont {Beserman}},\ }\href {\doibase
  10.1103/PhysRevB.61.3391} {\bibfield  {journal} {\bibinfo  {journal} {Phys.
  Rev. B}\ }\textbf {\bibinfo {volume} {61}},\ \bibinfo {pages} {3391}
  (\bibinfo {year} {2000})}\BibitemShut {NoStop}%
\bibitem [{\citenamefont {Gali}\ and\ \citenamefont {Maze}(2013)}]{gali13}%
  \BibitemOpen
  \bibfield  {author} {\bibinfo {author} {\bibfnamefont {A.}~\bibnamefont
  {Gali}}\ and\ \bibinfo {author} {\bibfnamefont {J.~R.}\ \bibnamefont
  {Maze}},\ }\href {\doibase 10.1103/PhysRevB.88.235205} {\bibfield  {journal}
  {\bibinfo  {journal} {Phys. Rev. B}\ }\textbf {\bibinfo {volume} {88}},\
  \bibinfo {pages} {235205} (\bibinfo {year} {2013})}\BibitemShut {NoStop}%
\bibitem [{\citenamefont {Schwoerer-B\"ohning}, \citenamefont {Macrander},\
  and\ \citenamefont {Arms}(1998)}]{schwoerer98}%
  \BibitemOpen
  \bibfield  {author} {\bibinfo {author} {\bibfnamefont {M.}~\bibnamefont
  {Schwoerer-B\"ohning}}, \bibinfo {author} {\bibfnamefont {A.~T.}\
  \bibnamefont {Macrander}}, \ and\ \bibinfo {author} {\bibfnamefont {D.~A.}\
  \bibnamefont {Arms}},\ }\href {\doibase 10.1103/PhysRevLett.80.5572}
  {\bibfield  {journal} {\bibinfo  {journal} {Phys. Rev. Lett.}\ }\textbf
  {\bibinfo {volume} {80}},\ \bibinfo {pages} {5572} (\bibinfo {year}
  {1998})}\BibitemShut {NoStop}%
\end{thebibliography}

%

\end{document}